\pdfoutput=1 
\pdfoutput=1 
\pdfoutput=1 
\pdfoutput=1 
\pdfoutput=1 
\documentclass[journal=jacsat,manuscript=article]{achemso}
\usepackage{tcolorbox}
\usepackage[version=3]{mhchem} 

\author{Javad Shabanpour}
\email{m.javadshabanpour1372@gmail.com}
\affiliation[Iran University of Science and Technology]
{Department of Electrical Engineering, Iran University of Science and Technology, Narmak, Tehran 16486-13114, Iran}

\title[An \textsf{achemso} demo]
{Fully manipulate the power intensity pattern in a large space-time digital metasurface: from arbitrary multibeam generation to harmonic beam steering scheme}

\abbreviations{IR,NMR,UV}
\begin{document}

\begin{abstract}
 Beyond the scope of space-coding metasurfaces, space-time digital metasurfaces can substantially expand the application scope of digital metamaterials in which simultaneous manipulation of electromagnetic waves in both space and frequency domains would be feasible. In this paper, by adopting a superposition operation of terms with unequal coefficient, Huygens principle, and a proper time-varying biasing mechanism, some useful closed-form formulas in the class of large digital metasurfaces were presented for predicting the absolute directivity of scatted beams. Moreover, in the harmonic beam steering scheme, by applying several suitable assumptions, we have derived two separate expressions for calculating the exact total radiated power at harmonic frequencies and total radiated power for scattered beams located at the end-fire direction. Despite the simplifying assumptions we have applied, we have proved that the provided formulas can still be a good and fast estimate for developing a large digital metasurface with a predetermined power intensity pattern. The effect of quantization level and metasurface dimensions on the performance of power manipulating as well as the limitation on the maximum scan angle in harmonic beam steering have been addressed. Several demonstrative examples numerically demonstrated
 through MATLAB software and the good agreement between
 simulations and theoretical predictions have been observed. By considering the introduced restrictions in the manuscript, this method can be implemented in any desired frequency just by employing phase-only meta-particles as physical coding elements. The author believes that the proposed straightforward approach discloses a new opportunity for various applications such as multiple-target radar systems and THz communication. 
\end{abstract}

\section{Introduction}
Artificial metamaterials and their 2D counterpart, called metasurfaces, have attracted widespread consideration due to their capabilities to tailor the permittivity
and permeability to reach values beyond material composites found in nature\cite{1,2}. Such metasurfaces which are immune to losses and easy to integrate can be structured for advanced manipulation of electromagnetic (EM) waves and have steadily witnessed significant growth in manipulating diverse
wave signatures such as phase, amplitude, and polarization\cite{3,4,11}.
Beyond the scope of analog metasurfaces, the
concept of digital metasurfaces has quickly evolved since they were initially introduced in 2014\cite{5}. 
This alternative approach for engineering the scattering patterns by designing two distinct coding elements with opposite reflection phases (e.g., 0° and 180°), has created a link
between the physical and digital worlds, making it possible to
revisit metamaterials from the perspective of information science\cite{6,7}.
However, in most of these strategies, the metasurfaces are designed for a specific application and their scattering program remains unchanged after being fabricated. Digital metasurfaces
accompanied by reprogrammable functionalities furnish 
a wider range of wave-matter functionalities which renders them especially appealing in the applications of imaging\cite{8}, smart surfaces\cite{9,37}, and dynamical THz wavefront manipulation\cite{10,35}.

All the above mentioned digital programmable architectures are space-coding metasurfaces wherein the 
coding sequences are generally fixed in time and are controlled through a computer-programmed biasing networks only to switch the functionalities whenever needed. To expand the application scope of digital metamaterials, the concept of space-time digital metasurface\cite{12} has been raised to obtain simultaneous manipulations of EM waves in both space and frequency domains in which the operational status of the
constituent meta-particles can be instantaneously controlled
through external digital time-domain signals. Spatiotemporal phase gradients provide additional degrees of freedom to control the normal momentum component, leading to a break in reciprocity during the light-matter interactions wherein such nonreciprocal effects can be controlled dynamically\cite{13}. Dai et al. experimentally characterized a time-domain digital
metasurface to manipulate the amplitude and phase for each harmonic independently\cite{14}.
Moreover, phase/amplitude modulation to implement a quadrature amplitude modulation (QAM) wireless communication is proposed and experimentally verified\cite{15}.

Fully manipulate the power intensity pattern of the metasurfaces can give us fabulous flexibility and privilege and is highly demanded in divers practical applications, such as direct broadcasting, multiple-target radar systems and MIMO communication\cite{16,17,18}. Formerly, a few studies have assisted
in addressing asymmetric multibeam reflectarrays producing
multiple beams with arbitrary beam directions and gain levels. Nayeri et al. proposed a single-feed reflectarray
with asymmetric multiple beams by implementing an
optimization process\cite{19}. However, this work has been realized
with a brute-force particle swarm optimization for producing
a phase profile of reflectarray elements resulting in a high computational cost that must be repeated afresh if the design characteristics change. Recently, based on the Huygens principle, by revisiting the addition theorem in the metasurface, we have
presented the concept of asymmetric spatial power divider
with arbitrary power ratio levels\cite{20,36}. Utilization  straightforward analytical methods and by modulating both amplitude and phase of the meta-atoms, one can estimate the directivity ratio levels of multiple beams. The above architecture suffers from two major drawbacks. Firstly, the proposed semi-analytical framework can predict the power level of each radiated beam but not the absolute value. Secondly, simultaneously modulate the amplitude and phase profiles of a metasurface necessitated utilizing the C-shaped meta-particles leading to call this structure as geometrically-encoded metasurface. Compared to the above work, wherein the lack
of predicting the absolute value of multibeam directivity as well as the lack of adjustability significantly hinders its practical applications, here, for the first time we present some useful approximation in a large space-time digital metasurface to predict the absolute directivity of each scattered multibeam with closed-form formulas. Therefore, both mentioned challenges will be resolved.

Accordingly, in this paper, by adopting superposition operation of terms with unequal coefficients on the electric field distribution and the Huygens principle, some convenient closed-form
formulation to predict the absolute directivity values for each
radiated multibeam is derived without any optimization procedure. Modulating both amplitude and phase of the meta-atoms is inevitable to fully manipulate the power intensity pattern of a metasurface. Besides, to control the power distribution in a reprogrammable manner, we have benefited from the concept of space-time digital metasurface where a set of coding sequences are switched cyclically in a predesigned time period. 
In the first section, the goal is to arbitrary manipulate the power distribution of large space-time metasurfaces which aimed to generate two beams in the desired directions with predetermined directivity values at the central frequency.
In this section, we introduced a time-varying biasing mechanism in which the summation of the total radiated power of all harmonics can be approximated as half of the total radiated power at the central frequency. 
In the second section, by considering a harmonic beam steering scheme in a large space-time digital metasurface, we have presented a set of closed-form formula to predict the exact value of directivity at any harmonic frequencies which shows amazing concordance with simulation results.   
This general concept can be implemented in any desired frequency just by employing phase-only meta-particles as physical coding elements. As a proof of concept, several illustrative examples
numerically demonstrated through MATLAB software. Eventually,
the simulated results have a very good agreement with
our theoretical prediction.
By designing phase-only meta-particles as a physical coding element and by encoding proper time-varying spatial codes, based on the presented formalism, our proposed structure can be implemented in space-time digital metasurface based systems\cite{31,32,33,34}. 
 Finally, at the end of the article, four investigations have been conducted to determine the limits of the validity range of the assumptions.
 The proposed straightforward approach is expected to broaden the applications of digital coding metasurfaces significantly and exposes a new opportunity for various applications such as multiple-target radar systems and NOMA communication\cite{21,22}. 
\section{1. Arbitrary multibeam generation with predetermined directivity at the central frequency}
We consider a space-time digital metasurface
that contains a square array of $N$×$N$ discrete elements characterized by a periodic time-coding sequence of length $L$ so that the digital layout of the proposed metasurfaces can then be demonstrated through a space-time-coding matrix as illustrated in \textbf{Fig. 1a}. According to the time-switched array theory\cite{23},  the Huygens’ principle\cite{20},  and approximations originating from the physical-optics, upon illuminating by a normal monochromatic plane wave, the far-field scattering pattern by the space-time digital metasurface for isotropic coding elements at the $m$th harmonic frequency can be expressed as\cite{12}:
\begin{equation}
{F^m}(\theta ,\varphi ,t) = \sum\limits_{q = 1}^N {\sum\limits_{p = 1}^N {\,a_{pq}^m\,\exp \left\{ {j\frac{{2\pi }}{{{\lambda _m}}}\left[ {(p - 1){d_x}\sin \theta \cos \varphi  + (q - 1){d_y}\sin \theta \sin \varphi } \right]} \right\}} } 
\end{equation}
where ${{d_x}}$ and ${{d_y}}$ are the
elements period along the $x$ and $y$ directions, respectively and ${\lambda _m} = c/({f_c} + m{f_0})$ is the wavelength of the reflected waves
corresponding to the $m$th harmonic frequency where the modulation frequency, ${f_0}$, is much smaller than the incident wave frequency, ${f_c}$\cite{24}.
${a_{pq}^m}$ is the Fourier series coefficients of time-modulated reflection
coefficient of the $(p,q)$th element and after some Fourier-based mathematical manipulations, one can deduce that:

\begin{equation}
\,a_{pq}^m = \sum\limits_{n = 1}^L {\frac{{\Gamma _{pq}^n}}{L}\,} {\rm{sinc}}\left( {\frac{{\pi m}}{L}} \right)\exp \left[ {\frac{{ - j\pi m(2n - 1)}}{L}} \right]
\end{equation}
where $\Gamma _{pq}^n = {\rm{A}}_{pq}^n\exp (j\varphi _{pq}^n)$ is the reflection coefficient of the $(p, q)$th coding element during the $n$th interval , i.e., $(n - 1){T_0}/L < t < n{T_0}/L$. Theoretically speaking, $a_{pq}^m$  specifies the equivalent amplitude and phase excitations
of all elements at a specific harmonic frequency. 

	\begin{figure*}[t]
	\centering
	\includegraphics[height=9cm]{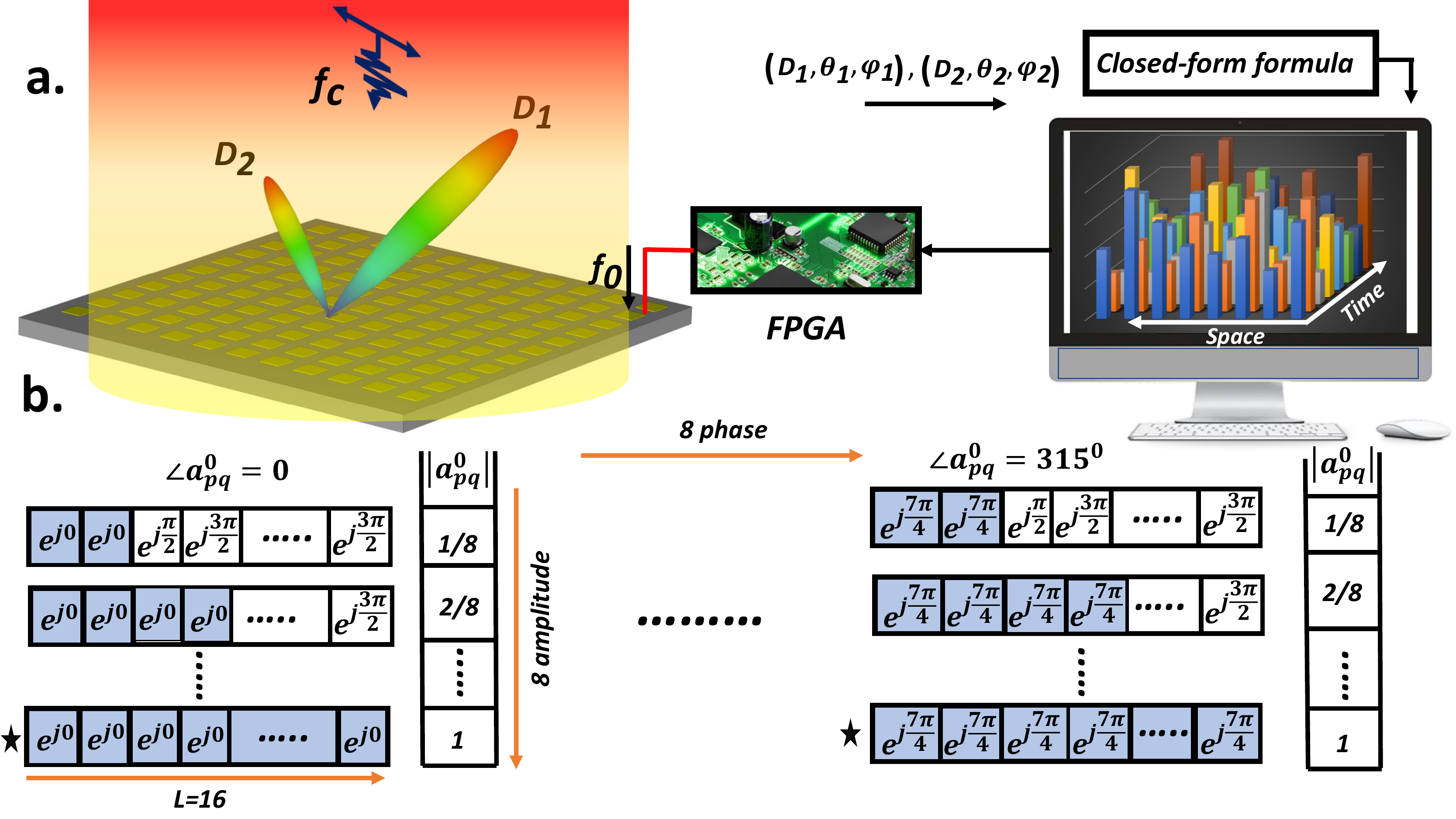}
	\caption{(a) Conceptual illustration of the proposed time-modulated metasurface aimed to divide the incident energy into two asymmetrically oriented beams with predetermined
		absolute directivity values at the central frequency. (b) Proposed time-varying biasing mechanism with 64 distinct phase/amplitude responses for eight level quantization.}
	\label{fgr:example2col}
\end{figure*}
As investigated in the previous work\cite{20}, fully control the power ratio levels necessitates implementing high quantization levels ($\ge$3-bit) for both amplitude and phase responses in which the building units of metasurface are characterized by 64 distinct phase/amplitude responses for eight-level quantization. Benefited from time modulated metasurface, independently modulate the amplitude and phase profiles of a
metasurface by adopting a phase-only meta-atoms has been realized as depicted in \textbf{Fig. 1b}. We suppose that the reflection amplitude ${\rm{A}}_{pq}^n$ of each meta-atom is uniform, while the reflection phase $\varphi _{pq}^n$ is a periodic function of time, whose values can be dynamically switched between eight different cases of [${0^\circ}$,${45^\circ}$,${90^\circ}$,${135^\circ}$,${180^\circ}$,${225^\circ}$,${270^\circ}$,${315^\circ}$]. Overall, $a_{pq}^m$ can be arbitrary tuned between 64 distinct phase/amplitude states. Each meta-atom has its own independent time-coding sequence, yielding various equivalent amplitudes and phases at the
separate harmonic frequencies. In the proposed time-varying biasing mechanism, the number of intervals should be considered $L = 2 \times \log _2^M$ in which, $M$ is the number of quantization bits. Accordingly, we set $L$=16 for eight level quantization (3-bit) in the first section of the paper (see \textbf{Fig. 1b}) 

Based on the superposition of the aperture fields, the
additive combination of two distinct phase-amplitude patterns
yields a mixed phase-amplitude distribution, whereby both individual functionalities will appear at the same time in the superimposed metasurface cause to reach a metasurface with several missions. We will demonstrate that by adding real-valued multiplicative constants, ${p_1}$ and ${p_2}$ into the conventional superposition operation, one can arbitrarily control the absolute value of directivity for each
multibeam independently through a closed form formula which is obtained by large metasurface assumption. In line with our outlined purpose,
we employ the superposition operation with unequal coefficients at the central frequency as follows:
\begin{equation}
{p_1}{e^{j{\phi _1}}} + {p_2}{e^{j{\phi _2}}} = \,\left| b \right|{e^{j{\phi _T}}}
\end{equation}
Here, $e^{j{\phi _T}}$ and $b$ carries  the phase and amplitude information of a superimposed metasurface respectively. To realize a multibeam metasurface at the central frequency, 
${e^{j{\phi _i}}}$ contains the pattern information of a single beam pointing at $({\theta _i},{\varphi _i})$ direction with uniform amplitude ($\left| {{e^{j{\phi _i}}}} \right| = 1$) and gradient phase distribution. After applying 3-bit quantization (64 distinct phase/amplitude responses) to $e^{j{\phi _T}}$ and $b$, the time-coding sequences of each individual coding elements will be obtained. Once the time coding sequences is determined according to time-varying biasing scheme presented in \textbf{Fig. 1b}, one can readily obtain the reflection coefficient of the $(p,q)$th element, $\Gamma _{pq}^n$ . Subsequently, based on Eq. 2, the equivalent phase and
amplitude levels of the meta-atoms at each harmonic, $a_{pq}^m$, will be calculated. \textbf{Fig. 2a} displays a schematic diagram of this process.
	\begin{figure*}[t]
	\centering
	\includegraphics[height=9cm]{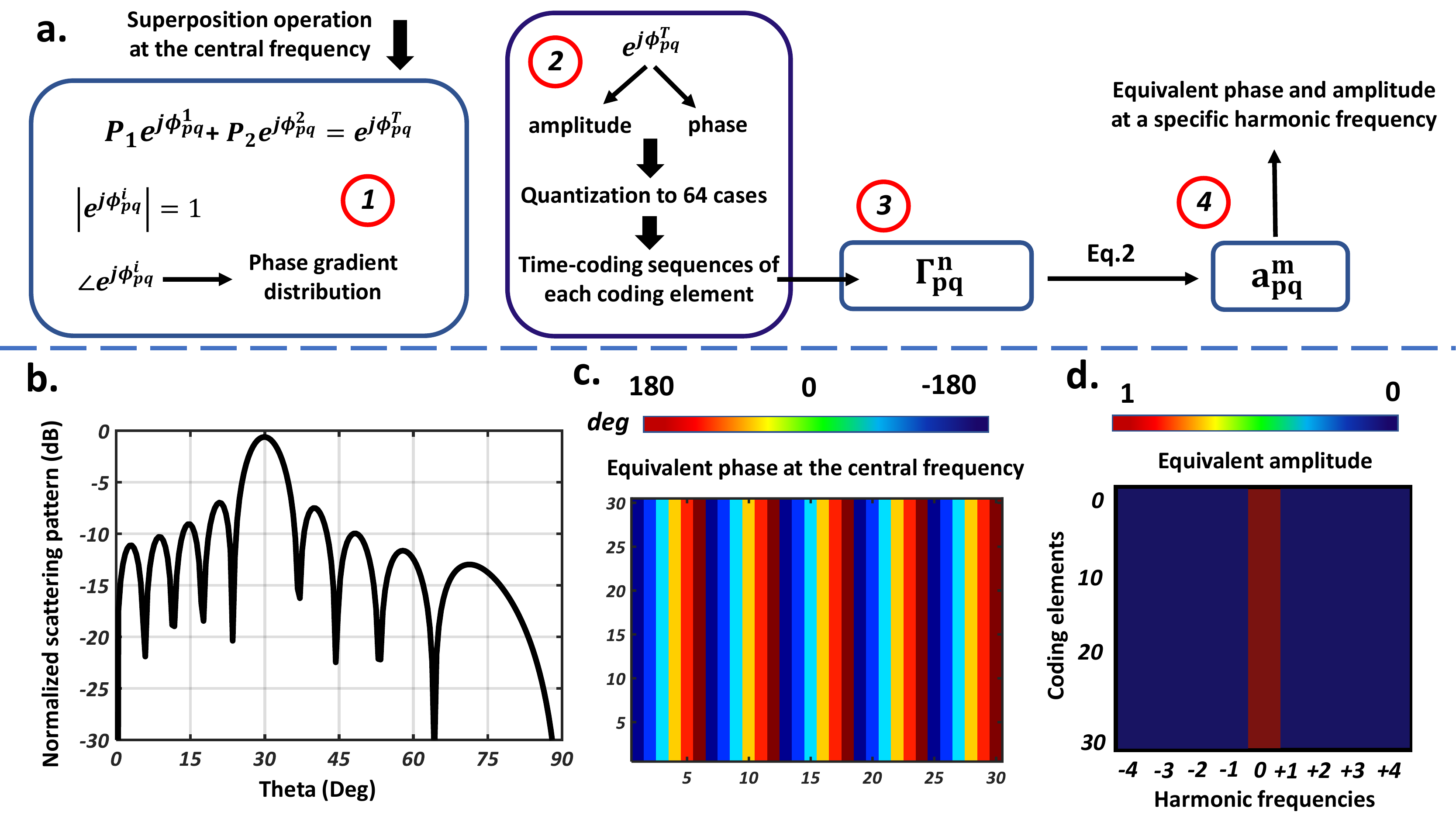}
	\caption{(a) Sketch representation of obtaining equivalent phase and amplitude at harmonic frequencies based on the superposition operation at the central frequency. (b) Normalized scattering pattern of a single-beam time-modulated metasurface pointing at $({30^o},{180^o})$ direction. (c) Corresponding equivalent phase at the central frequency. (d) Corresponding equivalent amplitude at different harmonic frequencies.}
	\label{fgr:example2col}
\end{figure*}

Thanks to the fact that the coding pattern and scattering pattern are a Fourier transform pair\cite{31}, by taking 2D IFFT from Eq. 3, then
   \begin{equation}
{p_1}F_0^1(\theta ,\varphi ) + {p_2}F_0^2(\theta ,\varphi ) = F_0^T(\theta ,\varphi )
\end{equation}
wherein $F_0^i(\theta ,\varphi )$ represents the array factor of primary metasurfaces and $F_0^T(\theta ,\varphi )$ stands for the superimposed array factor of the final two-beam metasurface at the central frequency.
To calculate the directivity of the space-time metasurface, the total radiated power contains all the Fourier components. Eventually, the peak directivity of the superimposed two-beam space-time metasurface can be computed by\cite{25}:
\begin{equation}
D(\theta ,\varphi ) = \frac{{4\pi {{\left| {{\rm{AF}}_0^T(\theta ,\varphi )} \right|}^2}_{\max }}}{{\sum\limits_{m =  - \infty }^\infty  {\int_0^{2\pi } {\int_0^\pi  {{{\left| {{\rm{AF}}_m^T(\theta ,\varphi )} \right|}^2}\sin \theta d\theta d\varphi } } } }}
\end{equation}
Then, the peak directivity of the radiated beam toward $({\theta _1},{\varphi _1})$ can
be calculated:
 \begin{equation}
D({\theta _1},{\varphi _1}) = \frac{{4\pi {{\left| {{\rm{AF}}_0^T({\theta _1},{\varphi _1})} \right|}^2}}}{{\int_0^{2\pi } {\int_0^\pi  {{{\left| {{\rm{AF}}_0^T(\theta ,\varphi )} \right|}^2}\sin \theta d\theta d\varphi } }  + {\mkern 1mu} 2\sum\limits_{m = 1}^\infty  {\int_0^{2\pi } {\int_0^\pi  {{{\left| {{\rm{AF}}_m^T(\theta ,\varphi )} \right|}^2}\sin \theta d\theta d\varphi } } } }}\, = \frac{{4\pi {{\left| {{\rm{AF}}_0^T({\theta _1},{\varphi _1})} \right|}^2}}}{{{{\rm{Q}}_1} + {{\rm{Q}}_2}}}
 \end{equation}
${{\rm{Q}}_2}$ represents the summation of the total radiated power of all harmonics and may not admit a general closed-form analytical
solution. But fortunately, according to the proposed biasing scheme, for different values of elevation angles and multiplicative constants in a two-beam space-time metasurface, ${{\rm{Q}}_2}$ can be approximated as ${{\rm{Q}}_2} \simeq \,0.5\,{{\rm{Q}}_1}$ (See section 1 in the  supporting information).  Since the superposition operation is adopted at the central frequency, for the sake of simplicity, we have defined $F_0^i(\theta ,\varphi ) = \,{F_i}(\theta ,\varphi )$ throughout this paper. 
By substituting Eq. 4 into Eq. 6 and applying above assumption, $D({\theta _1},{\varphi _1})$ becomes:
 \begin{equation}
D({\theta _1},{\varphi _1}) = \frac{{4\pi [{p_1}{F_1}({\theta _1},{\varphi _1}) + {p_2}{F_2}({\theta _1},{\varphi _1})][{p_1}F_1^*({\theta _1},{\varphi _1}) + {p_2}F_2^*({\theta _1},{\varphi _1})]}}{{1.5\int_0^{2\pi } {\int_0^{\pi /2} {[{p_1}{F_1}(\theta ,\varphi ) + {p_2}{F_2}(\theta ,\varphi )][{p_1}F_1^*(\theta ,\varphi ) + {p_2}F_2^*(\theta ,\varphi )]\sin \theta d\theta d\varphi } } }}
 \end{equation}
 In the above equation ${p_1}$ and ${p_2}$ are real-valued coefficients. For a large metasurface with negligible sidelobes, we suppose that  the angular position of the maximum in the array factor for the first beam is located in the vicinity of the null of the second beam, that is, ${F_2}({\theta _1},{\varphi _1}) \simeq 0$ and we can estimate the total radiated power as E2 presented in section 2 of the supporting information. It is worth noting that we can only employ the assumption of (E2) when we use an additive combination of distinct constant amplitude-gradient phase excitations to generate multibeam and the other methods to generate multibeam will encounter major errors. Although we apply these simplifying assumptions which will lead to closed-form formalism, we will show that they are valid with a very good approximation in an almost large digital metasurface. Furthermore, the limitation of the above assumptions has been addressed in section investigations (1)-(3).

 Numerical simulations are carried out for calculating the approximate and exact value of total radiated power at the central frequency (${{\rm{Q}}_1}$). The results can be found in section 2 of the supporting information. Applying above assumptions, Eq. 7 is simplified as:
 
     \begin{equation}
D({\theta _1},{\varphi _1}) = \frac{{4\pi p_1^2{{\left| {{F_1}({\theta _1},{\varphi _1})} \right|}^2}}}{{1.5(\int\limits_0^{2\pi } {\int\limits_0^{\pi /2} {p_1^2{{\left| {{F_1}(\theta ,\varphi )} \right|}^2}\sin \theta d\theta d\varphi  + \int\limits_0^{2\pi } {\int\limits_0^{\pi /2} {p_2^2{{\left| {{F_2}(\theta ,\varphi )} \right|}^2}\sin \theta d\theta d\varphi )} } } } }}\,
      \end{equation}
In the above equation ${p_1}$ and ${p_2}$ are real-valued coefficients. ${F_1}$ and ${F_1}$ represent the array factor of the first and second scattered beams at the center frequency.  
We use Jacobian for applying a variable change from $d\theta d\varphi $ to $d{\psi _x}d{\psi _y}$, ($F\left( {\theta ,\varphi } \right) \to F'\left( {{\psi _x},{\psi _y}} \right)$), in which, ${\psi _x} = 2\pi \frac{d}{\lambda }(\sin \theta \cos \varphi  - \sin {\theta _{\max }}\cos {\varphi _{\max }})$ and ${\psi _y} = 2\pi \frac{d}{\lambda }(\sin \theta \sin \varphi  - \sin {\theta _{\max }}\sin {\varphi _{\max }})$. ${\theta}$ and ${\varphi}$ are the
elevation and azimuth observation angles, respectively, $d$ indicates the periodicity of meta-atoms along both vertical
and horizontal directions and ${\lambda }$ is the working wavelength at the central frequency. ${\theta _{\max }}$ and ${\varphi _{\max }}$ represent the angles of maximum radiation with reference to broadside direction.
	\begin{equation}
d{\psi _x}d{\psi _y} = \left| {\begin{array}{*{20}{c}}
	{{\raise0.7ex\hbox{${\partial {\psi _x}}$} \!\mathord{\left/
				{\vphantom {{\partial {\psi _x}} {\partial \theta }}}\right.\kern-\nulldelimiterspace}
			\!\lower0.7ex\hbox{${\partial \theta }$}}}&{{\raise0.7ex\hbox{${\partial {\psi _x}}$} \!\mathord{\left/
				{\vphantom {{\partial {\psi _x}} {\partial \varphi }}}\right.\kern-\nulldelimiterspace}
			\!\lower0.7ex\hbox{${\partial \varphi }$}}}\\
	{{\raise0.7ex\hbox{${\partial {\psi _y}}$} \!\mathord{\left/
				{\vphantom {{\partial {\psi _y}} {\partial \theta }}}\right.\kern-\nulldelimiterspace}
			\!\lower0.7ex\hbox{${\partial \theta }$}}}&{{\raise0.7ex\hbox{${\partial {\psi _y}}$} \!\mathord{\left/
				{\vphantom {{\partial {\psi _y}} {\partial \varphi }}}\right.\kern-\nulldelimiterspace}
			\!\lower0.7ex\hbox{${\partial \varphi }$}}}
	\end{array}} \right| = {k^2}{d^2}\sin \theta \cos \theta d\theta d\varphi 
\end{equation}  
Since the metasurface is large and the beamwidth of each independent scattered beam is narrow, then, the major contributions to the integral of total radiated power for the first and second beam will be in the neighborhood of ${\theta _1}$ and ${\theta _2}$ respectively. 
Therefore, the expression of $\cos \theta $ which has appeared in the integral of total radiated power of the first and second beam can be approximated by $\cos {\theta _1}$ and $\cos {\theta _2}$ respectively\cite{27}.
In other words, the total radiated power for a single beam almost large metasurface along the $({\theta _i},{\varphi _i})$ direction can be written:   
	\begin{equation}
{P_{{\rm{radiation}}}}({\theta _i}) \cong \frac{1}{{\cos {\theta _i}}} \times {P_{{\rm{radiation}}}}({\rm{broadside}})
\end{equation} 
To verify the above equation, we have calculated the
${P_{{\rm{radiation}}}}$
for a scanned single beam metasurface for different values of scanning angles and metasurface length (See section 3 in the supporting information). As can be observed from Table S4, the comparison between approximate and exact results depict a perfect concordance. Besides, it can be concluded that Eq. 10 will be a very good approximation for metasurfaces with ${\rm{A}} > 5\lambda $. By substituting the above equation into Eq. 8, $D({\theta _1},{\varphi _1})$ becomes:
	\begin{equation}
D({\theta _1},{\varphi _1}) \cong \frac{{4\pi {k^2}{d^2}p_1^2{{\left| {F'(0,0)} \right|}^2}}}{{1.5\left( {\frac{{p_1^2}}{{\cos {\theta _1}}} + \frac{{p_2^2}}{{\cos {\theta _2}}}} \right)\left( {\int {\int_\Omega  {{{\left| {F'({\psi _x},{\psi _y})} \right|}^2}d{\psi _x}d{\psi _y}} } } \right)}},\,\,\Omega  = {({\psi _x})^2} + {({\psi _y})^2} \le {k^2}{d^2}
\end{equation} 
In the above equation, $F'({\psi _x},{\psi _y})$ stands for the array factor in a broadside direction and has a uniform excitation amplitude ($\left| {{e^{j{\phi _i}}}} \right| = 1$) and would be as the product of those
two linear arrays \cite{28}, then 
	\begin{equation}
F'({\psi _x},{\psi _y}) = F'({\psi _x})F'({\psi _y})
\end{equation}
By applying Eq. 12 which is known as separable or multiplication method, the rest of calculation can be found as follows:
	\begin{equation}
D({\theta _1},{\varphi _1}) \cong \frac{{\pi p_1^2}}{{1.5\left( {\frac{{p_1^2}}{{\cos {\theta _1}}} + \frac{{p_2^2}}{{\cos {\theta _2}}}} \right)}} \times \frac{{2kd{{\left| {F'(0)} \right|}^2}}}{{\int\limits_{ - kd}^{kd} {{{\left| {F'({\psi _x})} \right|}^2}d{\psi _x}} }} \times \frac{{2kd{{\left| {F'(0)} \right|}^2}}}{{\int\limits_{ - kd}^{kd} {{{\left| {F'({\psi _y})} \right|}^2}d{\psi _y}} }}
\end{equation}
\vspace{5pt}
	\begin{equation}
D({\theta _1},{\varphi _1}) \cong \frac{{\pi p_1^2}}{{1.5\left( {\frac{{p_1^2}}{{\cos {\theta _1}}} + \frac{{p_2^2}}{{\cos {\theta _2}}}} \right)}}{D_x}{D_y} = \frac{{\pi p_1^2}}{{1.5\left( {\frac{{p_1^2}}{{\cos {\theta _1}}} + \frac{{p_2^2}}{{\cos {\theta _2}}}} \right)}} \times \frac{{2{\rm{A}}}}{\lambda } \times \frac{{2{\rm{A}}}}{\lambda }
\end{equation}
\vspace{5pt}
\begin{tcolorbox}
	\begin{equation}
D({\theta _1},{\varphi _1}) = \frac{{\frac{2}{3}\cos {\theta _1}}}{{1 + {{\left( {\frac{{{p_2}}}{{{p_1}}}} \right)}^2}\left( {\frac{{\cos {\theta _1}}}{{\cos {\theta _2}}}} \right)}} \times {D_{\max }}
\end{equation}
\end{tcolorbox}
\vspace{5pt}
\begin{tcolorbox}
	\begin{equation}
D({\theta _2},{\varphi _2}) = \frac{{\frac{2}{3}{{\left( {\frac{{{p_2}}}{{{p_1}}}} \right)}^2}\cos {\theta _1}}}{{1 + {{\left( {\frac{{{p_2}}}{{{p_1}}}} \right)}^2}\left( {\frac{{\cos {\theta _1}}}{{\cos {\theta _2}}}} \right)}} \times {D_{\max }}
	\end{equation}
\end{tcolorbox}

It should be noted that, in deducing Eq. 14, ${D_x}$ and ${D_y}$ represent the peak directivity of linear arrays along $x$ and $y$ directions and equal to ${{2A} \mathord{\left/
		{\vphantom {{2A} \lambda }} \right.
		\kern-\nulldelimiterspace} \lambda }$ in which $A$ denotes the length of the array. 
Following the same steps, the peak directivity of a two-beam space-time metasurface with proposed time-varying biasing mechanism toward $({\theta _2},{\varphi _2})$ can be immediately obtained from Eq. 16.	
	Overall, the absolute value of directivity along $({\theta _1},{\varphi _1})$ and $({\theta _2},{\varphi _2})$ direction can be immediately obtained from Eq. 15 and Eq. 16 respectively where ${D_{\max }}$ represents the maximum directivity  of a metasurface and equals to ${{4\pi {{\rm{A}}^2}} \mathord{\left/
		{\vphantom {{4\pi {{\rm{A}}^2}} {{\lambda ^2}}}} \right.
		\kern-\nulldelimiterspace} {{\lambda ^2}}}$ (${\rm{A}} = Nd$).
The variable $N$ denotes the number of meta-atoms in a proposed space-time metasurface and can be selected differently for desired value of directivities as below:
	\vspace{12pt}	 
\begin{tcolorbox}
	\begin{equation}
N = \frac{\lambda }{d}\sqrt {\frac{3}{{8\pi }}\left( {\frac{{D({\theta _1},{\varphi _1})}}{{\cos {\theta _1}}} + \frac{{D({\theta _2},{\varphi _2})}}{{\cos {\theta _2}}}} \right)} 
	\end{equation}
\end{tcolorbox}
It should be noted that for a space-time metasurface generating a single beam at desired direction (${p_2} = 0$), it is required that all the meta-atoms have the phase gradient distribution with uniform reflection amplitude. In this case, the time-coding sequences  will be obtained in such a way that  the reflection phase of each individual coding elements are constant during each 16 intervals which is marked with stars in \textbf{Fig. 1b}. Then, the equivalent amplitude at harmonic frequencies will be equal to zero (See \textbf{Fig. 2b}). 
\begin{equation}
\left| {a_{pq}^m} \right| = \sum\limits_{n = 1}^L {\frac{{{e^{jX}}}}{L}\,} {\rm{sinc}}\left( {\frac{{\pi m}}{L}} \right)\exp \left[ {\frac{{ - j\pi m(2n - 1)}}{L}} \right] = 0
\end{equation}
$X$ is constant for each coding elements in a modulation period and can take any arbitrary value from 0 to ${{7\pi } \mathord{\left/
		{\vphantom {{7\pi } 4}} \right.
		\kern-\nulldelimiterspace} 4}$ .
	Therefore, the total radiated power at harmonics will be equal to zero $({{\rm{Q}}_2} = 0)$. Following the previous steps (Eq. 6-15), the peak directivity of a single beam metasurface along $({\theta _1},{\varphi _1})$ direction is equal to $D = \pi {D_x}{D_y}\cos {\theta _1}$. This is the well-known Elliott's expression for directivity of large scanning planar array\cite{29}.
	 
	\textbf{Concept Verification}. In order to demonstrate the fully manipulate the power intensity pattern, we will introduce two approaches to design
	a large space-time metasurface to generate two arbitrarily oriented reflected beams with predetermined  absolute directivity values. According to Eq. 17, when the dimensions of the metasurface are constant (${D_{\max }}$ and $N$ are fixed), by arbitrarily determining the directivity of
	the first beam, the directivity value of the second beam is inevitably determined. In the latter approach, the dimensions of
	the metasurface are considered as unknown and by arbitrarily determining the directivity of two beams, the length of the metasurface (A) or the number of coding elements ($N$) can be immediately obtained from Eq. 17. The numerical simulations
	are carried out in the MATLAB software employing the well-known antenna array theory. Normally incident plane-wave illumination is considered. Without loss of generality, the inter-element
	spacing between coding elements is considered ${d_x} = {d_y} = {\lambda  \mathord{\left/
			{\vphantom {\lambda  {3}}} \right.
			\kern-\nulldelimiterspace} {3}}$ in all the simulations.
For the sake of simplicity, we have defined ${D_i} = D({\theta _i},{\varphi _i})$ throughout the manuscript. In line with the first approach, the number of coding elements is fixed to $N=$ 30 which lead to ${D_{\max }}=$ 31dBi.

In the following, we will present an illustrative example in which the space-time metasurface divides the reflected energy between two multiple beams oriented along $({15^o},{180^o})$ and 
$({35^o},{270^o})$ directions. We applied conventional superposition operation (${p_1} = {p_2} = 1$) and the directivity of these two beams will be equal to ${D_1} = {D_2}$ = 25.7 dBi based on Eq. 15,16. Referring to the Huygens principle, numerical simulations are
performed for such an encoded space-time digital metasurface and the simulated 1D directivity intensity pattern is depicted in \textbf{Fig. 3a}. Outstandingly, the analytical predictions based on the Huygens principle and the superposition theorem estimates well the absolute directivity of the first and second beams as 25.74 dBi (less than 0.2\%
error). The quantitative comparison between simulation results and theoretical predictions are depicted in \textbf{Fig. 3d}.
 \textbf{Fig. 3c} shows the equivalent phase and amplitude at the central frequency. \textbf{Fig. 3b} illustrates the normalized total radiated power for each harmonic from -12th to +12th harmonic frequencies. As can be observed, benefited from the presented time-varying biasing mechanism, ${{\rm{Q}}_2}$ can be approximated as ${{\rm{Q}}_2} \simeq \,0.5\,{{\rm{Q}}_1}$.
 
	\begin{figure}[t]
	\centering
	\includegraphics[height=9cm]{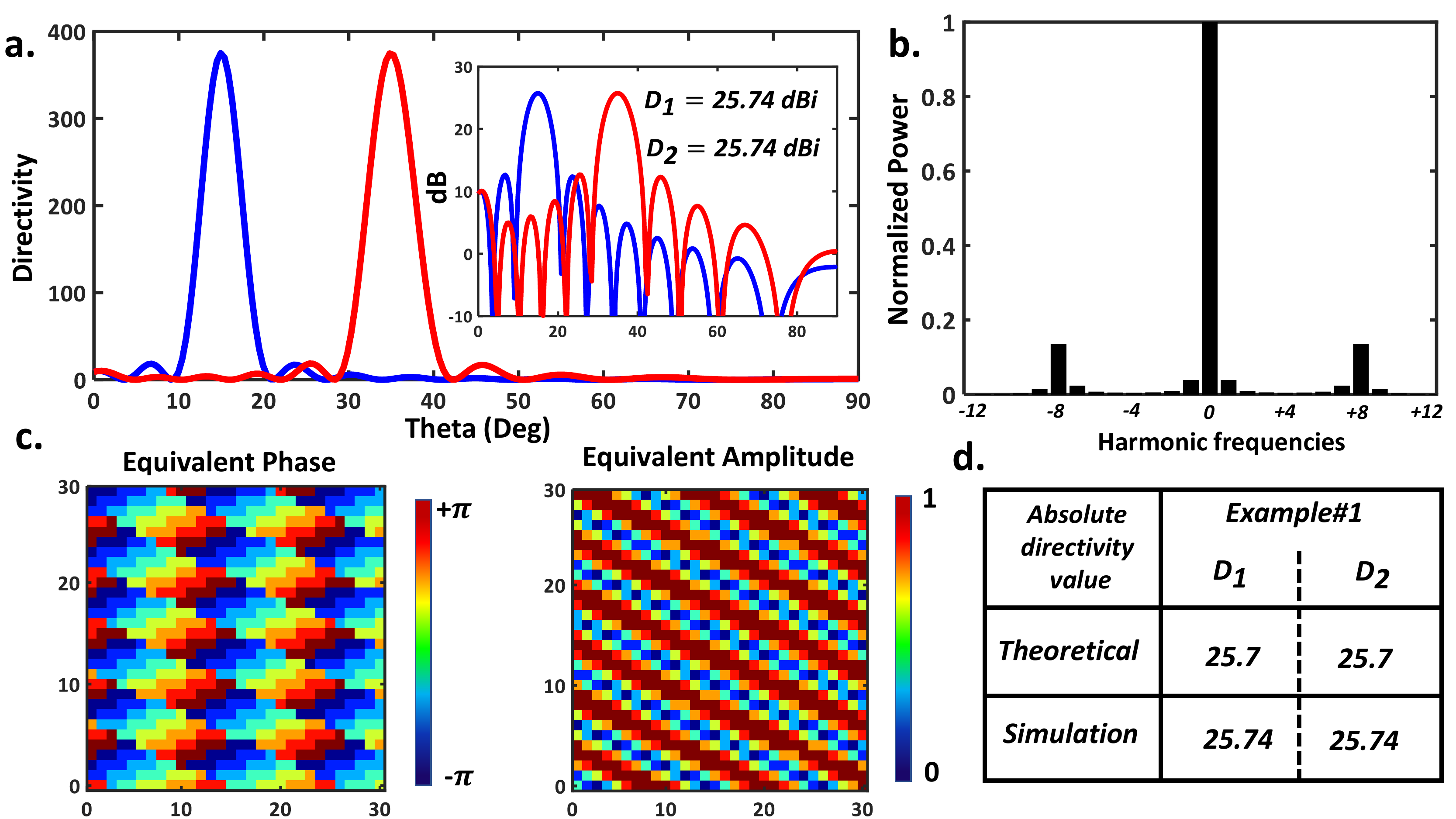}
	\caption{(a) Directivity intensity pattern (in both linear and decibel formats) for a space-time metasurface that scattered the incident wave into two beams with equal directivity values along $({15^o},{180^o})$ and 
		$({35^o},{270^o})$. (b) Normalized total radiated power for each harmonic frequency. (c) Corresponding equivalent phase and amplitude at the central frequency. (d) Comparison between simulation results and theoretical predictions.}
	\label{fgr:example2col}
\end{figure}

As a new scenario, we intend to design a two-beam
generating space-time digital metasurface along $({15^o},{180^o})$ and 
$({40^o},{270^o})$ directions whose absolute directivity along $({15^o},{180^o})$ direction is ${D_1}=$ 25 dBi. Since the maximum
directivity is constant, the absolute directivity along $({40^o},{270^o})$ direction is determined as follows which is equals to ${D_2}=$ 25.91 dBi 
\begin{equation}
\left( {\frac{{{D_1}}}{{\cos {\theta _1}}} + \frac{{{D_2}}}{{\cos {\theta _2}}}} \right) = \frac{2}{3}{D_{\max }}
 \end{equation}	
The real-valued multiplicative constants can be immediately obtained by dividing Eq. 15 into Eq. 16 and will be equal to ${p_1}=$ 0.9 and ${p_2=}$ 1 respectively (${{{p_1}} \mathord{\left/
		{\vphantom {{{p_1}} {{p_2} = }}} \right.
		\kern-\nulldelimiterspace} {{p_2} = }}$ 0.9).	
\begin{equation}
\frac{{{p_1}}}{{{p_2}}} = \sqrt {\frac{{{D_1}}}{{{D_2}}}} 
 \end{equation}		
After calculating the amplitude and phase pattern of the superimposed metasurface ($b$ and ${e^{j{\phi _T}}}$) from Eq. 3 and applying eight-level quantization, the time coding sequences of each coding element will be obtained according to proposed time-varying biasing scheme. Eventually, by performing numerical simulations, such an encoded metasurface plays the role of a large space-time digital metasurface architecture that
elaborately splits the normal incident wave into two asymmetric beams with the  directivity values of ${D_1}=$  24.98 (0.02 dB difference) and ${D_2}=$  26 dBi (0.09 dB difference). As can be observed in \textbf{Fig. 4},
the absolute directivity of two scattered beams satisfactorily
approaches the predetermined values with the desired tilt angles. The quantitative comparison of the aforesaid results is tabulated in \textbf{Fig. 4d}.
Consequently,
one can conclude that weighted combination of individual phase-only patterns in the framework of the superposition operation with unequal coefficient and the Huygens principle will
significantly boost the speed of designing the multiple beams
space-time metasurface just by employing phase-only meta-particles as physical coding elements.
	\begin{figure*}
	\centering
	\includegraphics[height=8cm]{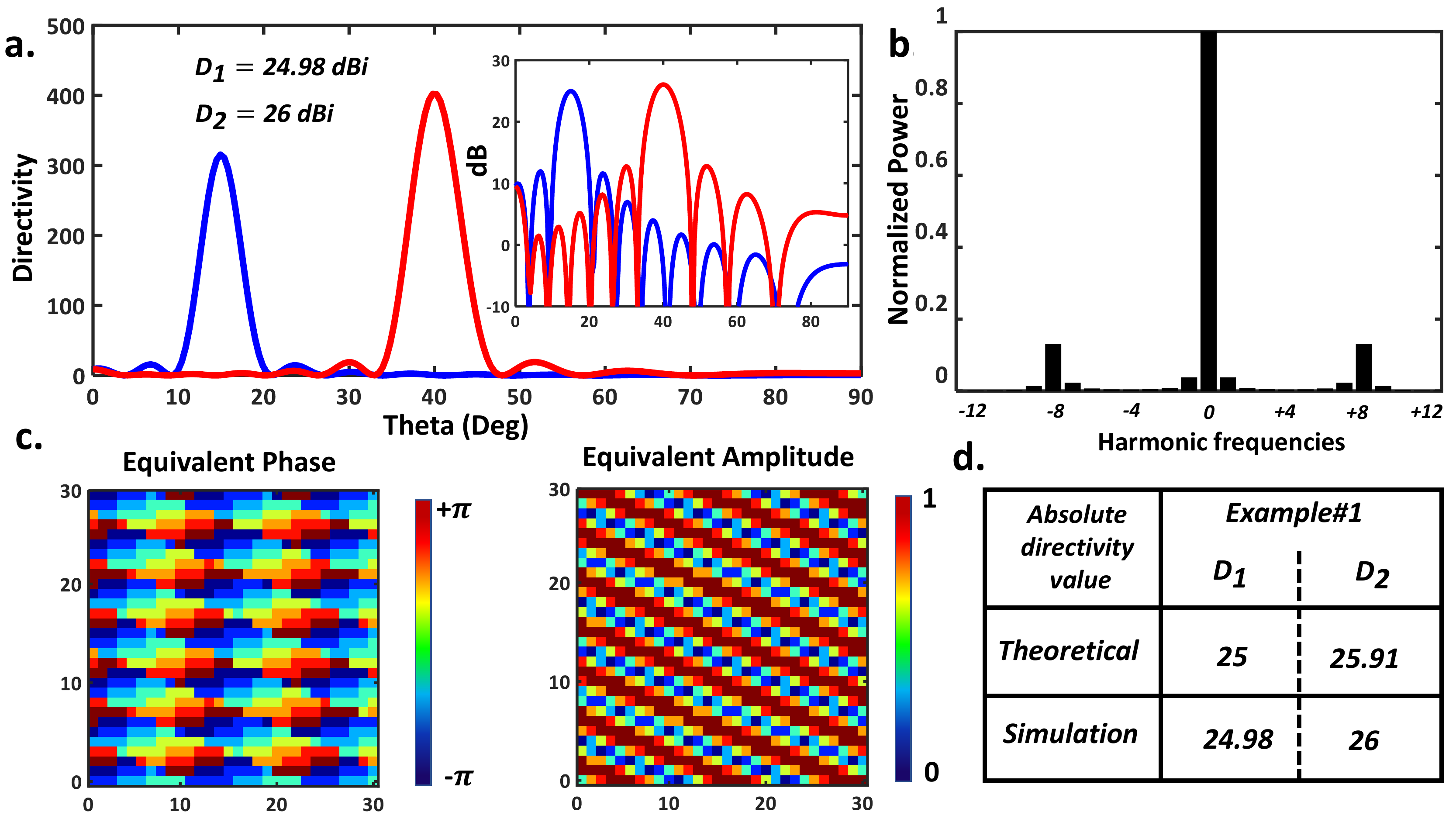}
	\caption{(a) Directivity intensity pattern (in both linear and decibel formats) for asymmetrically oriented two-beams time-modulated metasurface pointing at $({15^o},{180^o})$ and 
		$({40^o},{270^o})$. (b) Normalized total radiated power for each harmonic frequency. (c) Corresponding equivalent phase and amplitude at the central frequency. (d) Comparison between simulation results and theoretical predictions.}
	\label{fgr:example2col}
\end{figure*}

In line with the second approach, we consider the number
of coding elements unknown and by arbitrary determining the
absolute directivity of two scattered beams, one can immediately calculate the number of coding elements ($N$) based on the closed-form formulation presented in Eq. 17. As a new scheme, we intend to design
a space-time metasurface to generate two independent scattered beams pointing at $({18^{\circ}},{180^{\circ}})$ and $({32^{\circ}},{270^{\circ}})$. We wish the proposed digital metasurface to deflect the incident plane wave into two asymmetric reflected beams with ${D_1}=$  25.11 dBi and ${D_2}=$  23.72 dBi. Referring to Eq. 17, then the number of coding elements is obtained $N=$ 26. Based on the Huygens principle and the general form of the superposition theorem in Eq. 3,
the space-time metasurface must be endowed by the superimposed phase/amplitude
pattern obtained by assuming (${p_1}=$ 1, ${p_2}=$ 0.85) and $N=$ 26 to expose two asymmetrically 
oriented beams with predetermined directivities. As can be seen in \textbf{Fig. 5a}, the directivity value of two scattered beams satisfactorily approaches to ${D_1}=$ 25.11, ${D_2}=$ 23.69 dBi, 
that is very close to our theoretical predictions (See \textbf{Fig. 5d}).
The existing very negligible discrepancies can be attributed to
the nature of approximations applied to reach the closed-form
formulation. The equivalent phase and amplitude at the central frequency and the total normalized radiated power of the -12th to +12 harmonic frequencies are depicted in \textbf{Fig. 5c} and \textbf{Fig. 5b} respectively. 
	\begin{figure*}[t]
	\centering
	\includegraphics[height=9cm]{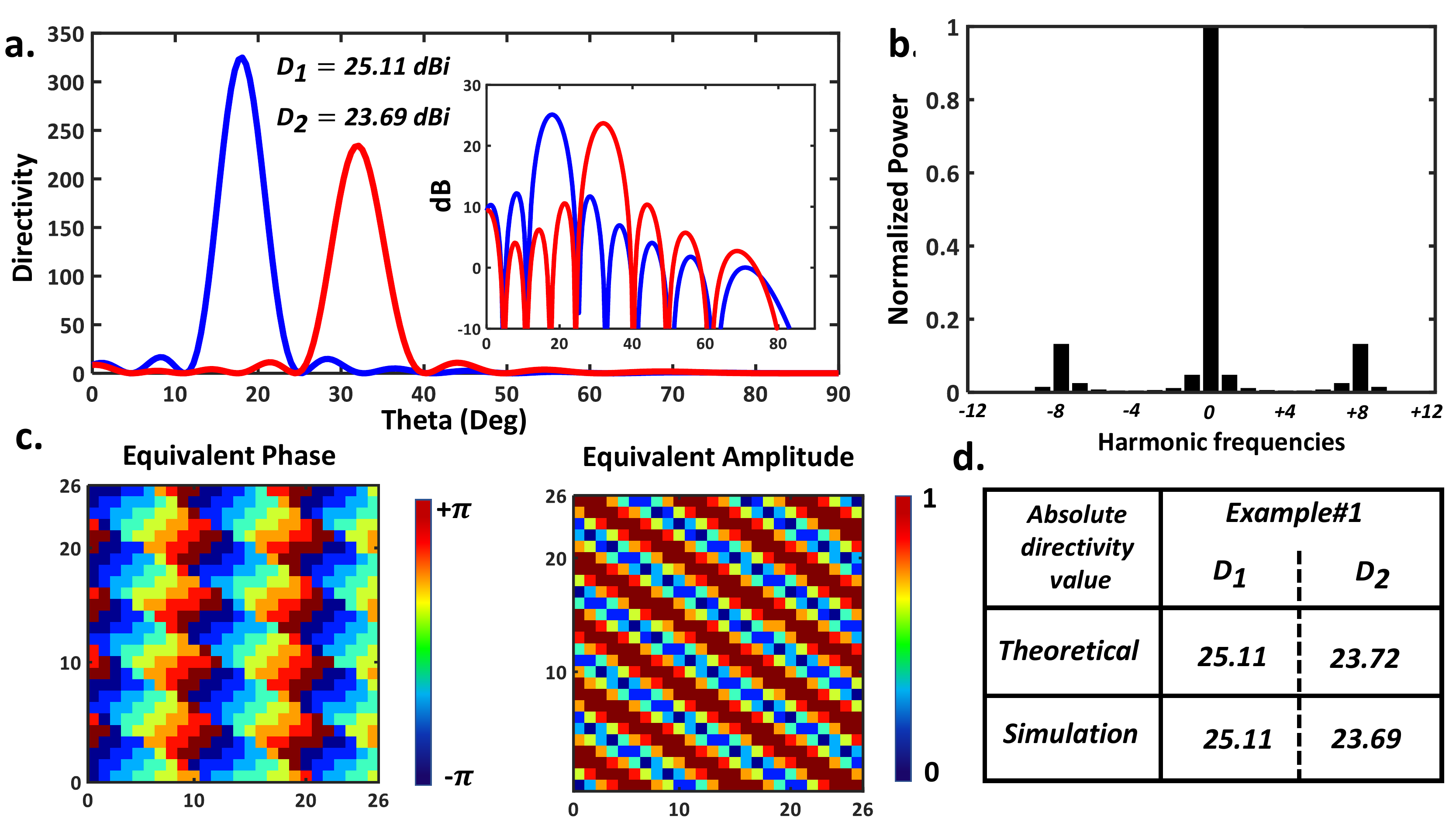}
	\caption{(a) Directivity intensity pattern (in both linear and decibel formats) for asymmetrically oriented two-beams time-modulated metasurface pointing at $({18^o},{180^o})$ and 
		$({32^o},{270^o})$. (b) Normalized total radiated power for each harmonic frequency. (c) Corresponding equivalent phase and amplitude at the central frequency. (d) Comparison between simulation results and theoretical predictions.}
	\label{fgr:example2col}
\end{figure*}

To further verify the concept and dive into the performance
of proposed method, our final example is devoted to a two-beam generating space-time digital metasurface with desired tilt angles pointing at $({15^{\circ}},{270^{\circ}})$ and $({65^{\circ}},{180^{\circ}})$, with predetermined directivities of ${D_1}=$ 25 and ${D_3}=$ 26.32 dBi. Referring to Eq. 17 and Eq. 20, the number of coding elements is equals to $N=$ 38 and the values of real-valued multiplicative constants become ${p_1}=$ 0.88, ${p_2}=$ 1 respectively. Referring to the directivity intensity pattern presented in \textbf{Fig. 6}, thanks to the 
superposition operation and Huygens principle, the space-time digital metasurface driven by the proper
phase-amplitude pattern obtained by Eq. 3 divides the incident energy into two asymmetrically oriented beams with ${D_1}=$ 25.06 dBi (0.02 dB difference) and ${D_2}=$ 26.29 dBi (0.3 dB difference). As ${\theta _i} \to {90^o}$, the expression in Eq. 10 is no longer valid due to the nature of the approximation. In section investigation (1), we have provided a limit for the validity of the above equations. In this example, the second beam has a large scan angle and must be checked whether it has exceeded the limit. Since the dimension of the metasurface in the proposed example is $10\lambda  \times 10\lambda$ , the limit will be equal to $70.5^\circ$ which is higher than the scan angle of the second beam ($65^\circ$).   
 
Overall,
the presented approach founded on closed-form formulation
successfully performs its missions, that is, predicting the absolute directivities of multiple beams which also
furnish an inspiring platform for realizing a space-time digital metasurface with
predetermined directivities pointing at desired directions without resorting to any brute-force optimization schemes. As can be deduced from the above examples, despite the simplifying assumptions we have applied, the provided formula can still be a good and quick estimate for designing a large digital metasurface. Exploring the limit of the metasurface dimensions for the accuracy of the above formulas is also discussed in section investigation (2).
\begin{figure*}
	\centering
	\includegraphics[height=9cm]{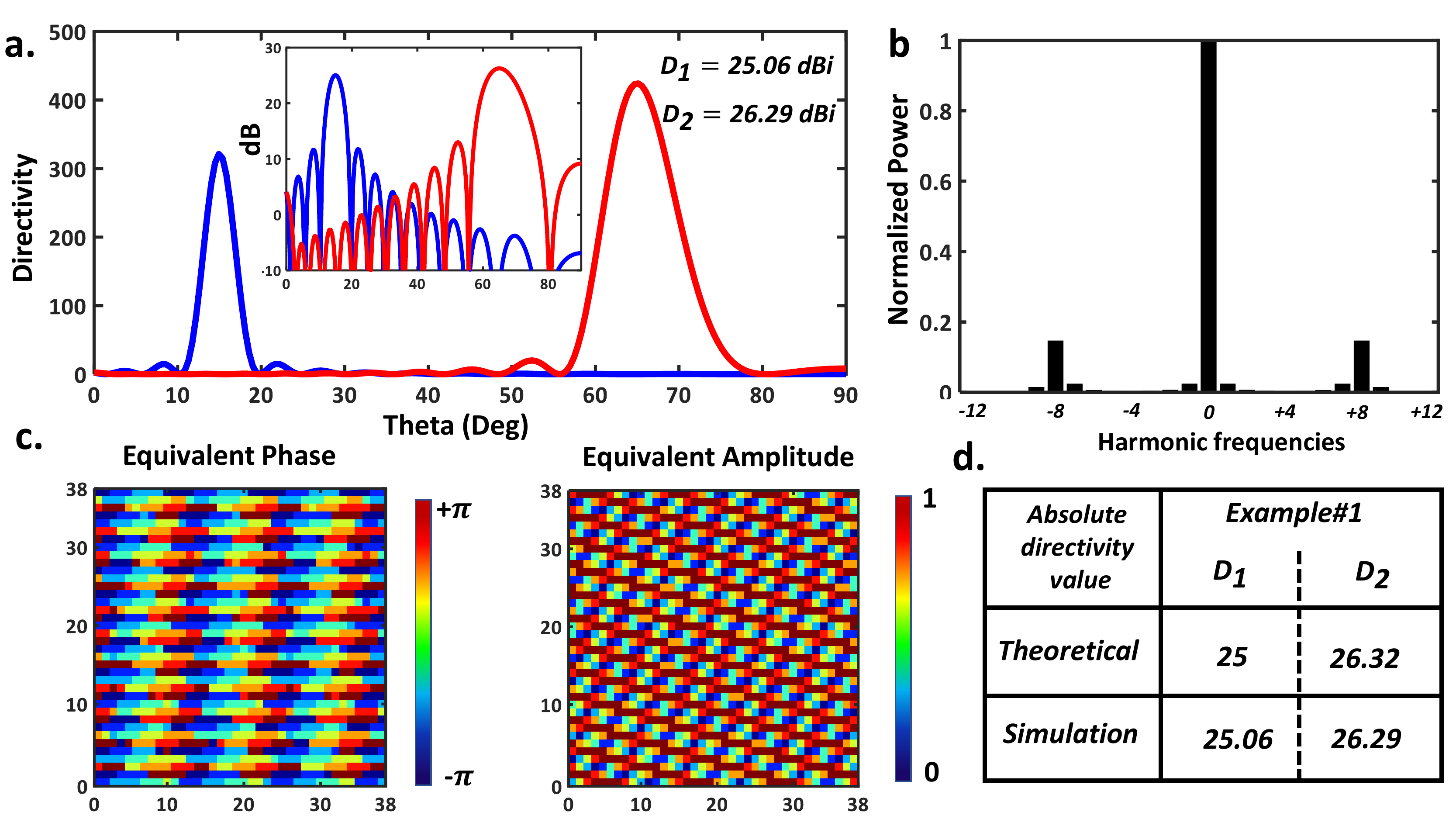}
	\caption{(a) Directivity intensity pattern (in both linear and decibel formats) for asymmetrically oriented two-beams time-modulated metasurface pointing at $({15^o},{180^o})$ and 
		$({65^o},{270^o})$. (b) Normalized total radiated power for each harmonic frequency. (c) Corresponding equivalent phase and amplitude at the central frequency. (d) Comparison between simulation results and theoretical predictions.}
	\label{fgr:example2col}
\end{figure*} 
\section{2. Directivity of harmonic beam steering}
In this section, we intend to present a closed-form formula to estimate the exact value of absolute directivity at any harmonic frequencies based on the proposed PM  harmonic beam steering scheme which is presented in Ref 12. The required number of coding elements to reach the predetermined directivity is also discussed. Unlike the previous section where we applied the assumption of orthogonality, in harmonic beam steering scheme each scattered beam is orthogonal to each other and the closed-form formulas obtained in this section are more rigorous.
  
PM harmonic beam steering can be realized by using time-gradient sequences as illustrated in \textbf{Fig. S3} in which the phase $\varphi _{pq}^n$ is a periodic function of time whose values are either ${0^\circ}$ or ${180^\circ}$. Again, the directivity function of such an encoded space-time metasurface can be computed by Eq. 5 in which the total radiated power contains all the Fourier components. According to the proposed time-varying biasing scheme (See \textbf{Fig. S3} in the supporting information), the equivalent amplitude at the central and harmonic frequencies are constant for all the coding elements and can be obtained as follows respectively.
\begin{equation}
\left| {a_{pq}^0} \right| = \sum\limits_{n = 1}^L {\frac{{{e^{jY}}}}{L}\,}  = \,\frac{{L - 2}}{L}
\end{equation}
\begin{equation}
\left| {a_{pq}^m} \right| = \sum\limits_{n = 1}^L {\frac{{{e^{jY}}}}{L}{\mkern 1mu} } {\rm{sinc}}\left( {\frac{{m\pi }}{L}} \right)\exp \left[ {\frac{{ - j\pi m(2n - 1)}}{L}} \right] = \frac{2}{L}{\mkern 1mu} {\rm{sinc}}\left( {\frac{{m\pi }}{L}} \right)
\end{equation}

In $L$ intervals, for each coding element, Y consists of one "1" digits and $L-1$ "0" digits. In the harmonic beam steering scheme,  the beampattern at the fundamental frequency (${f_c}$) always has a maximum response at the boresight direction ($\theta  = 0$). The phase difference ($\Delta {\psi _m}$) between
neighboring coding elements at the $m$th harmonic frequency can be written as:
\begin{equation}
\Delta {\psi _m} = \frac{{2m\pi }}{L}
\end{equation}
Under normal illumination, the scattered beam at the $m$th harmonic frequency obeys the generalized snell's law and the beam steering angle ${\theta _m}$ can be written as:
\begin{equation}
{\theta _m} = \arcsin \left( {\frac{{\Delta {\psi _m}}}{{2\pi {d_\lambda }}}} \right) = \arcsin \left( {\frac{m}{{L{d_\lambda }}}} \right)
\end{equation}
In the above equation ${d_\lambda } = {{{d_y}} \mathord{\left/
		{\vphantom {{{d_y}} \lambda }} \right.
		\kern-\nulldelimiterspace} \lambda }$, in which ${{d_y}}$ is the inter-element spacing between meta-atoms in the phase progressive direction ( in this paper along $y$-direction). From the knowledge gained in the previous section (Eq. 10) and Eq .21-22 in this section, one can deduce that:  
\begin{equation}	
{P_m} = {\left[ {\frac{2}{{L - 2}} \times {\rm{sinc}}\left( {\frac{{m\pi }}{L}} \right)} \right]^2}\,\frac{{{P_0}}}{{\cos \left( {{\theta _m}} \right)}}	
\end{equation}
In which ${{P_0}}$ and ${{P_m}}$ are the total radiated power at the fundamental and harmonic frequencies respectively in a large space-time digital metasurface.
As ${\theta _m} \to {90^o}$, this expression is no longer valid due to the
nature of the approximation in Eq. 10. According to Eq. 24, the beampatterns at the harmonic frequencies of $m = (2n - 1)L{d_\lambda }$, $n = 1,2,...$, have a maximum response at endfire direction. Referring to the method adopted by King and Thomas\cite{30},
at endfire, the total radiated power can be written as:
\begin{equation}
{P_{{\rm{endfire}}}} = \frac{{\frac{{4\pi {A^2}}}{{{\lambda ^2}}}}}{{\frac{{3\pi A}}{\lambda }\sqrt {{\raise0.7ex\hbox{${2A}$} \!\mathord{\left/
					{\vphantom {{2A} \lambda }}\right.\kern-\nulldelimiterspace}
				\!\lower0.7ex\hbox{$\lambda $}}} }}\,{P_{{\rm{broadside}}}} = \frac{4}{3}\sqrt {\frac{A}{{2\lambda }}} \,{P_{{\rm{broadside}}}}
\end{equation}
where $A$ is the length of the metasurface and equals to $A = Nd$. It should be noted that, since the phase difference between adjacent coding elements at $m = (2n - 1)L{d_\lambda }$  is $180^\circ$, then the metasurface divides the incident energy into two symmetrically oriented scattered beams. Therefore, for a large space-time digital metasurface, the total radiated power at the endfire direction is twice the value obtained in the above equation and it can be written with respect to total radiated power at the central frequency as follows:
\begin{equation}
{P_{m = (2n - 1)L{d_\lambda }}} = {\left[ {\frac{2}{{L - 2}} \times \,{\rm{sinc}}\left( {\frac{{m\pi }}{L}} \right)} \right]^2} \times \frac{8}{3}\sqrt {\frac{A}{{2\lambda }}} {P_0}
\end{equation}
Using Eq .25 and Eq. 27, one can estimate the exact total radiated power at all harmonic frequencies. Eventually, since the beampatterns at different harmonic frequencies are orthogonal to each other, the total radiated power in the harmonic beam steering scheme can be readily obtained:
\begin{tcolorbox}
\begin{equation}
\begin{array}{l}
{P_{total}} = \left[ {1 + 2{{\left( {\frac{2}{{L - 2}}} \right)}^2}\left[ {{R_1} + {R_2}} \right]} \right]{P_0}\\
\\
{R_1} = \sum\limits_{m = 1}^\infty  {\frac{{{\rm{sin}}{{\rm{c}}^2}\left( {\frac{{m\pi }}{L}} \right)}}{{\cos {\theta _m}}}} \,\,\,,\,m \ne (2n - 1)L{d_\lambda },\\
\\
{R_2} = \sum\limits_{m = (2n - 1)L{d_\lambda }}^\infty  {\frac{8}{3}\sqrt {\frac{A}{{2\lambda }}} \,\,{\rm{sin}}{{\rm{c}}^2}\left( {\frac{{m\pi }}{L}} \right)} 
\end{array}
\end{equation}
\end{tcolorbox}
To survey the validation of Eq. 25, 27, 28, numerical simulations are carried out employing the well-known antenna array theory. The coding metasurface composed of 40$\times$40 elements with 20-interval periodic time modulation ($L = 20$) and ${d_x} = {d_y} = {\lambda  \mathord{\left/
		{\vphantom {\lambda  2}} \right.
		\kern-\nulldelimiterspace} 2}$. 
 The total radiated power at the central frequency is equal to ${P_0} = 5256.2$. Quantitative comparison between the numerical simulations and theoretical prediction from 1st to 50th positive harmonic frequencies is illustrated in \textbf{Fig. 8}. As can be deduced from the phase difference between adjacent coding elements (Eq. 23), the beampattern at harmonic frequencies of m=1, 19, 21, 39, and 41 have the same steering elevation angle (See \textbf{Fig. 7a}).
	\begin{figure}
	\centering
	\includegraphics[height=10cm]{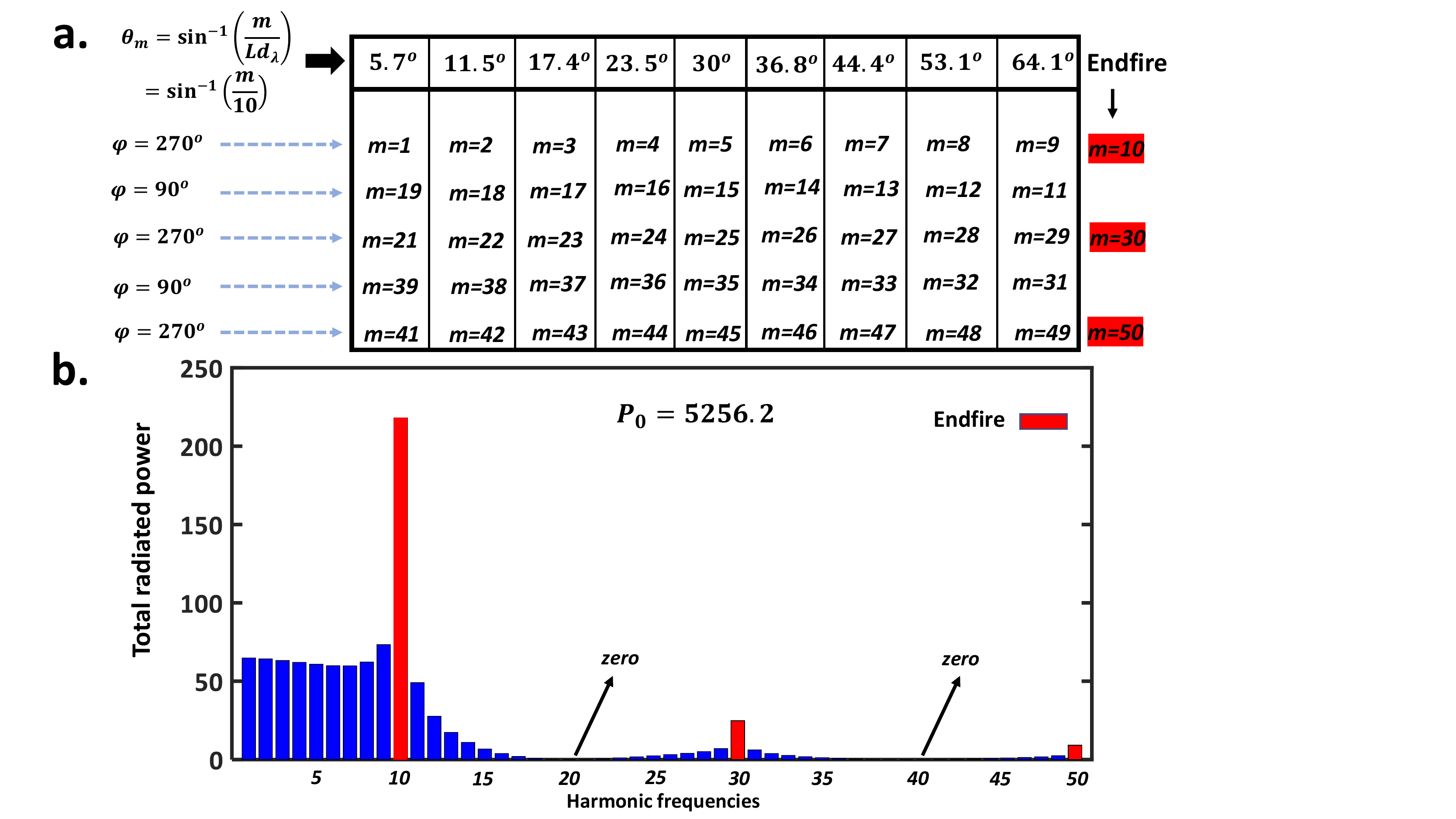}
	\caption{(a) The steering elevation and azimuth angles at each harmonic frequency. (b) The value of total radiated power at each harmonic from 1st to 50th positive harmonic frequencies when the total radiated power at the central frequency is equal to 5256.2.}
	\label{fgr:example2col}
\end{figure}
	\begin{figure}
	\centering
	\includegraphics[height=8cm]{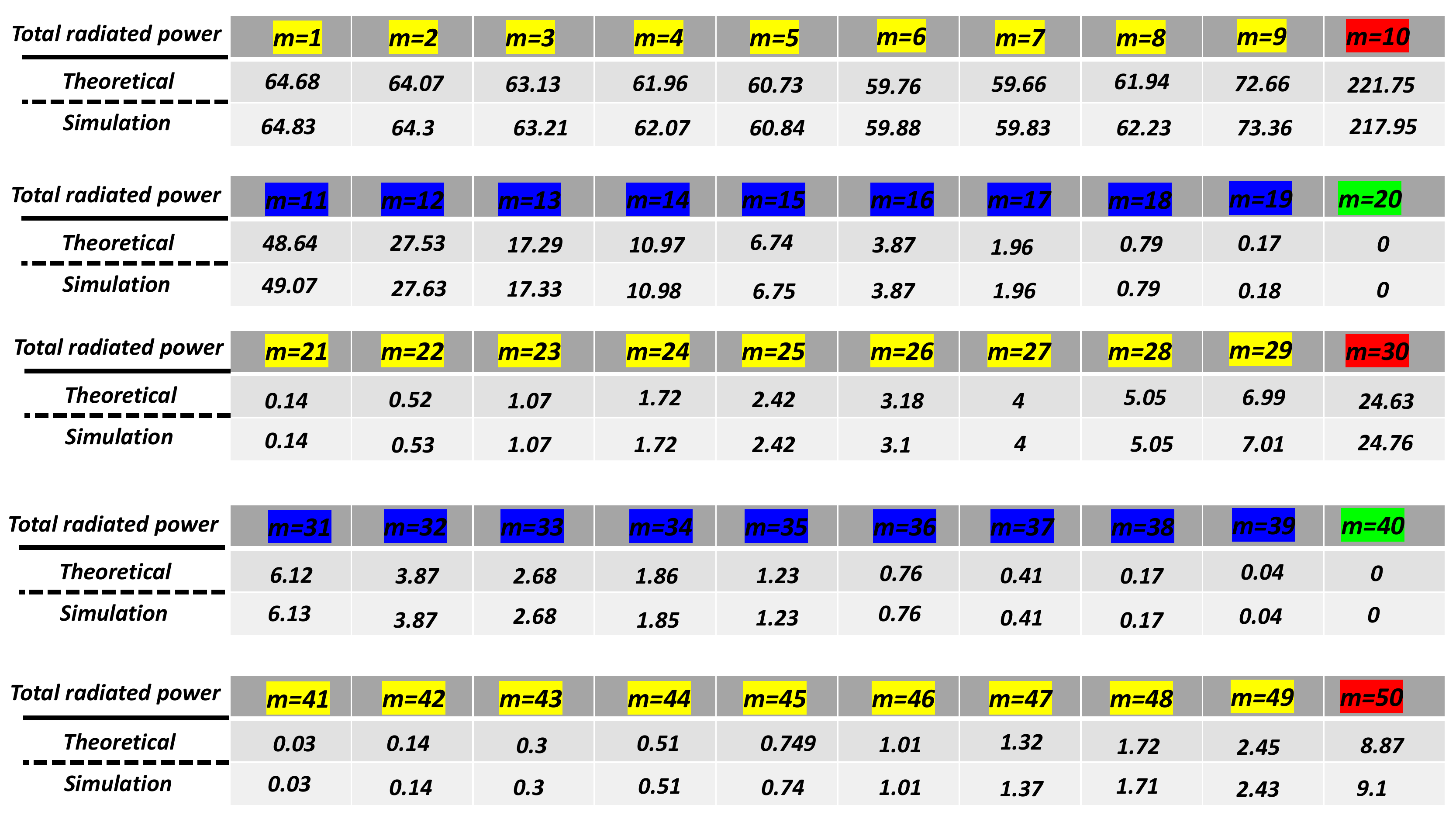}
	\caption{Quantitative comparison between simulation results and theoretical predictions for total radiated power at each harmonic from 1st to 50th positive harmonic frequencies.}
	\label{fgr:example2col}
\end{figure} 
As depicted in \textbf{Fig. 7b}, the beam patterns which are highlighted in red are located at the end-fire direction and the corresponding total radiated power can be estimated using Eq. 27, while the total radiated power of the other harmonics (blue bars in \textbf{Fig. 7b}) can be predicted using Eq. 25. The quantitative comparison between simulation results and theoretical predictions which are tabulated in \textbf{Fig. 8} shows amazing concordance. The harmonics highlighted in yellow and blue represent the scattering beams with azimuth angles of $\varphi  = 270^\circ$ and $\varphi  = 90^\circ$ respectively, while those highlighted in red have the maximum response at the end-fire direction. Although the harmonics of $m$=20 and $m$=40 are located in the broadside direction but their corresponding equivalent amplitude and total radiated power are zero.
\newpage
The absolute directivity of space-time metasurface in the harmonic beam steering scheme at the central and harmonic frequencies can be readily obtained respectively:
	\vspace{12pt}
\begin{tcolorbox}
\begin{equation}
{D_0} = \frac{{{D_{\max }}}}{{1 + 2{{\left( {\frac{2}{{L - 2}}} \right)}^2}\left[ {{R_1} + {R_2}} \right]}}
\end{equation} 
\end{tcolorbox}

\begin{tcolorbox}
\begin{equation}
{D_m} = \frac{{{{\left[ {\frac{2}{{L - 2}} \times {\rm{sinc}}\left( {\frac{{m\pi }}{L}} \right)} \right]}^2}}}{{1 + 2{{\left( {\frac{2}{{L - 2}}} \right)}^2}\left[ {{R_1} + {R_2}} \right]}} \times \,{D_{\max }}
\end{equation} 
\end{tcolorbox}
In the above equations, ${R_1}$ and ${R_2}$ are introduced in Eq. 28 and ${D_{\max }}$ represents the maximum directivity of the metasurface and is equal to ${{4\pi {A^2}} \mathord{\left/
		{\vphantom {{4\pi {A^2}} {{\lambda ^2}}}} \right.
		\kern-\nulldelimiterspace} {{\lambda ^2}}}$.
Despite the number of harmonics is unlimited, but for higher harmonics, the equivalent amplitude and corresponding total radiated power drop sharply. Therefore, the numerical simulations are carried out to calculate the peak directivity of each scattered beams in which the total radiated power of the space-time metasurface is calculated by considering -50th to +50th harmonic frequencies.	
\textbf{Fig. 9a} illustrates the simulated 1D directivity intensity pattern (linear format) at different harmonic frequencies (1st to 9th). Quantitative comparison between numerical simulations
and theoretical predictions are also detailed in \textbf{Fig. 9b}.	Outstandingly, the analytical predictions estimate well the absolute directivity and beam scanning angles and very negligible discrepancies can be attributed to the nature of approximations applied to reach the closed-form formulation, which is interestingly less than 2\%.
Based on theoretical simulations, according to \textbf{Fig. 8}, in our proposed example, 37\% of the incident energy is converted into high-order harmonics. The efficiency is obtained from the energy ratio between the harmonic and incident wave. This ratio is calculated by considering -50th to +50th harmonic frequencies. The majority of the energy is assigned to the first positive and negative harmonics located at the end-fire direction, as expected by Eq. 28. 
For almost large space-time digital metasurface, we can estimate the number of coding elements to reach the predetermined directivity of desired harmonics as follows:

	\begin{equation}
N = \frac{\lambda }{{d\left( {\frac{2}{{L - 2}} \times {\rm{sinc}}\left( {\frac{{m\pi }}{L}} \right)} \right)}}\, \times \,\sqrt {\frac{{{D_m}\left( {1 + 2{{\left( {\frac{2}{{L - 2}}} \right)}^2}\left[ {{R_1} + {R_2}} \right]} \right)}}{{4\pi }}} 
\end{equation} 
This is a good approximation in a large space-time metasurface which significantly boosts the speed of designing the harmonic beam steering scheme without resorting to any brute-force optimization.
	\begin{figure}
	\centering
	\includegraphics[height=9cm]{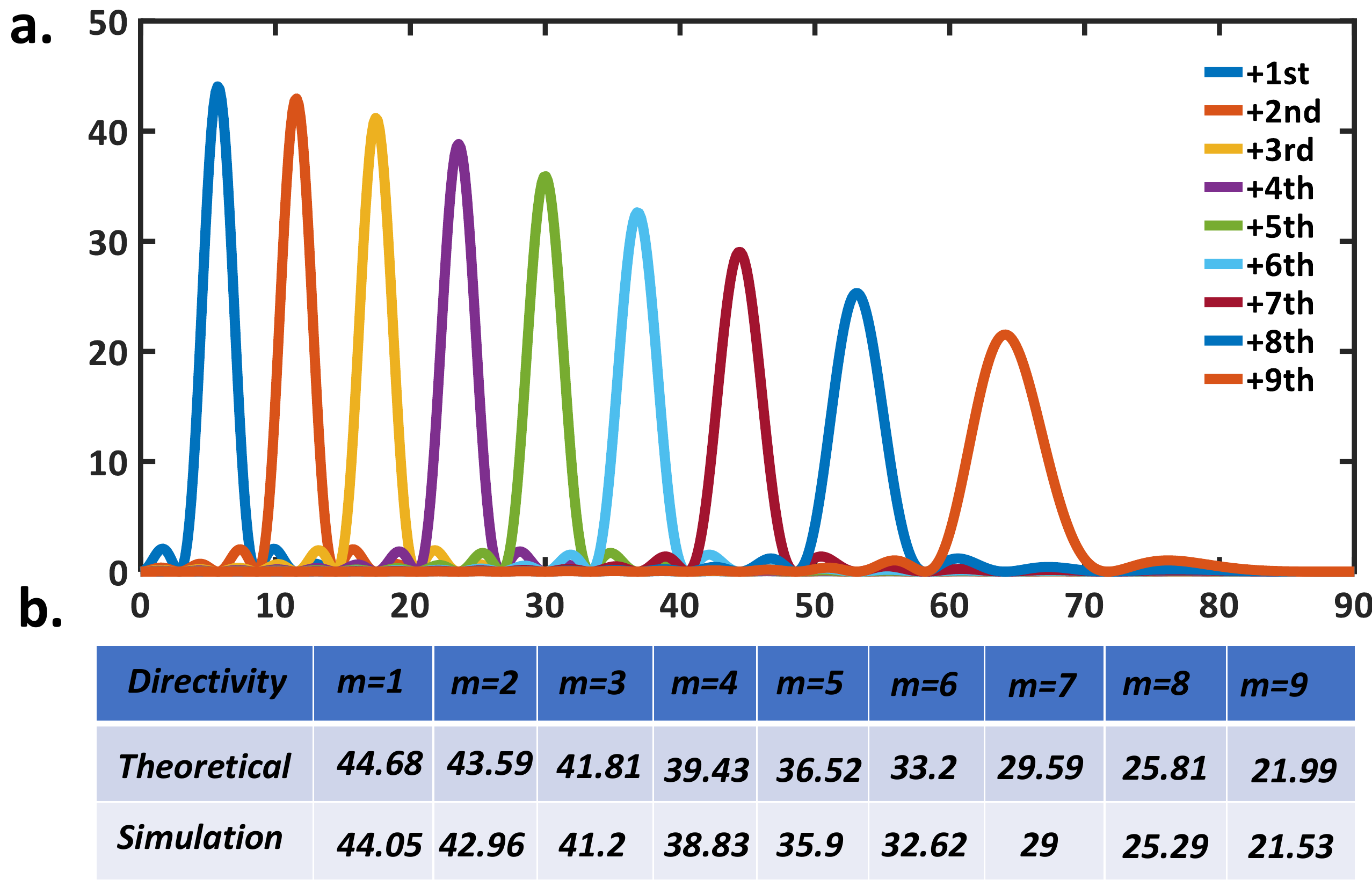}
	\caption{(a) 1D directivity intensity pattern (linear format) at different harmonic frequencies from 1st to 9th. (b) Quantitative comparison between numerical simulations and theoretical predictions.}
	\label{fgr:example2col}
\end{figure} 
\subsection{Investigation (1)}
Approximation provided in Eq. 10 and Eq. 25 is not valid for large scan angles. Therefore, we have introduced a new approximation for harmonics $m = (2n - 1)L{d_\lambda }$, which is presented in Equation 27. But the scan angle for the harmonics $m = L{d_\lambda } - 1$ ($m$=9, 19, 29,... in this paper) is still large and must be checked whether it has exceeded the limitation which is presented in Ref 30. By equating the two expressions for the directivity of large scanning array and directivity of endfire array, we have: 	
	\begin{equation}	
\frac{{4\pi {A^2}\cos {\theta _m}}}{{{\lambda ^2}}} = \,3\pi \frac{A}{\lambda }\sqrt {\frac{{2A}}{\lambda }} 
\end{equation}
Therefore, Eq. 25 is valid for the harmonics with scan angle lower than that given by the limiting case as expressed by Eq. 32.
	\begin{equation}	
{\theta _m} \le \,{\cos ^{ - 1}}\sqrt {\frac{{9\lambda }}{{8A}}} 
\end{equation}
In the presented example, the scan angle of the harmonic $m = L{d_\lambda } - 1$ is equal to $64.1^\circ$ (see Fig. 7a) which is lower than the limiting case obtained by Eq. 33 which is equal to $76.2^\circ$. 	
\subsection{Investigation (2)}
In this investigation, we will survey the impact of the metasurface dimensions in validation of Eq. 10. As mentioned before, if the beamwidth of scattered beams is narrow, the major contributions to the integral of the total radiated power will be in the neighborhood of the maximum scan angle. To further clarify, several numerical simulations are performed which are depicted in \textbf{Fig. 10}. In all simulations, the metasurfaces are encoded with the superimposed phase-amplitude pattern, $\left| b \right|\,{e^{j{\phi _T}}}$, obtained by assuming ${p_1} = 1.2$ and ${p_2} = 1$ to expose two differently oriented beams pointing at $({10^{\circ}},{180^{\circ}})$ and $({50^{\circ}},{270^{\circ}})$ directions. As can be noticed in \textbf{Fig. 10}, when $A < 5\lambda$,
the absolute directivity of the scattered beams does not further match with our theoretical predictions, thereby, the significant role of the metasurface length in validating Eq. 10 is highlighted. It should be noted that by decreasing the length of the metasurface, the limitation for maximum scan angle which is introduced in investigation (1) will be more restricted, thereby for $A = 5\lambda$, $8\lambda $, and $10\lambda $, this limitation would be $61.6^\circ$, $68^\circ$, and $70.5^\circ$ respectively.  
\begin{figure}[t]
\centering
\includegraphics[height=9cm]{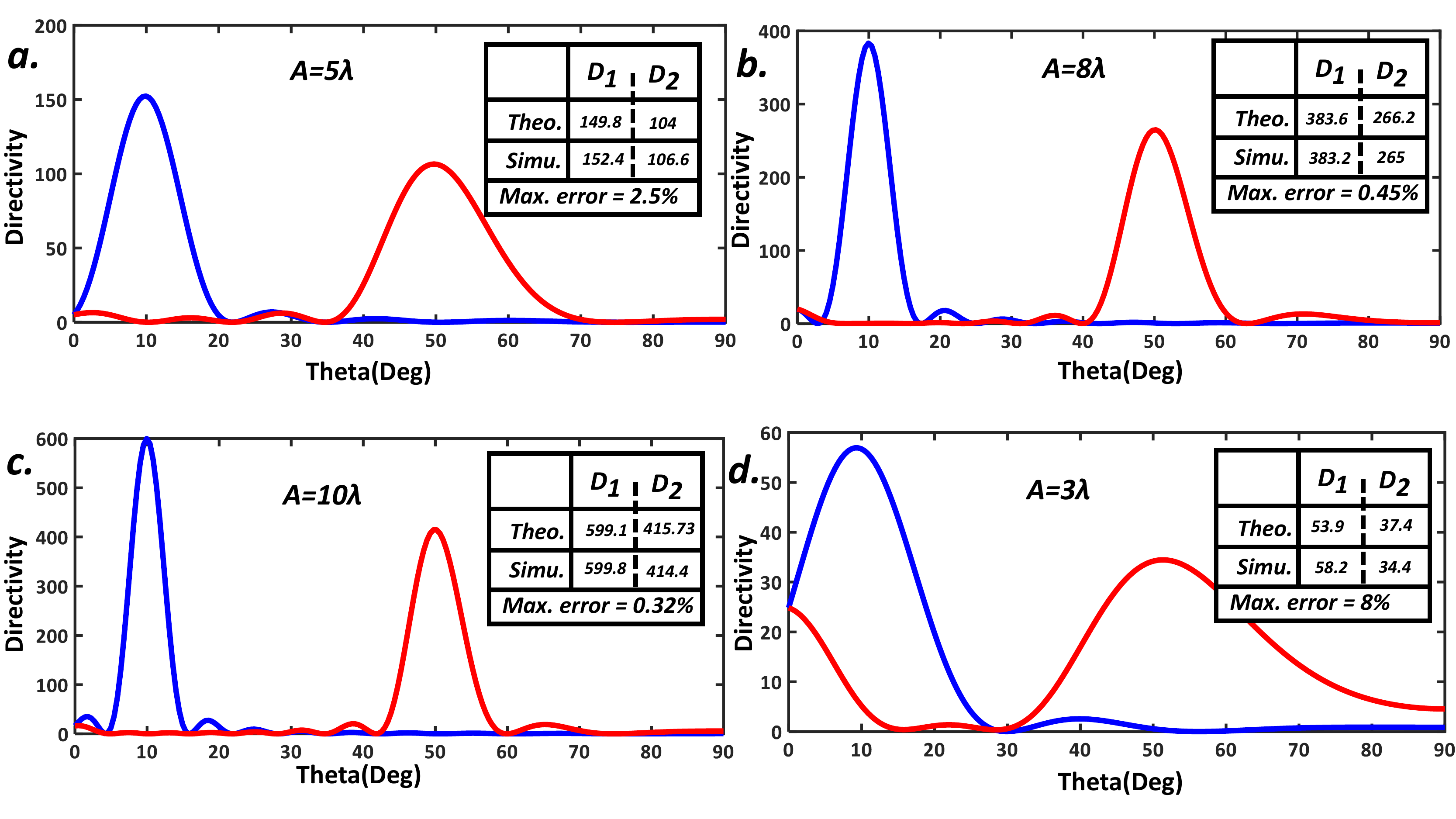}
\caption{1D directivity intensity pattern (linear format) for space-coding metasurface with asymmetric beams toward $({10^{\circ}},{180^{\circ}})$ and $({50^{\circ}},{270^{\circ}})$ directions when the length of the metasurface is equal to (a) $5\lambda$. (b) $8\lambda$. (c) $10\lambda$. (d) $3\lambda$.}
\label{fgr:example2col}
\end{figure} 

\subsection{Investigation (3)}
In the first section, we applied 3-bit quantization to the superimposed phase-amplitude pattern and we claimed that fully manipulate the power intensity pattern necessitates implementing at least 3-bit quantization level. In investigation (3), we will demonstrate how aggressive 
quantization deteriorates the performance of the closed-form formula presented in section 1. To investigate the quantization impact, the numerical simulations have been accomplished where the phase-amplitude profiles describing the superimposed metasurfaces are quantized into two or three levels. A fair comparison between the power intensity
patterns generated by the continuous and quantized
phase-amplitude profiles have been performed in \textbf{Fig. 11a-b}.
In the first illustration, the metasurface with
two-level (\textbf{Fig. 11a}) and three-level quantization (\textbf{Fig. 11b})
serve to divide the incident power between two scattered beams oriented along$({15^{\circ}},{180^{\circ}})$ and $({45^{\circ}},{270^{\circ}})$ directions with ${p_1} = 1$ and ${p_2} = 0.8$. In the next example, the metasurface with two-level (\textbf{Fig. 11c}) and three-level quantization (\textbf{Fig. 11d}) sets for ${p_1} = 1$ and ${p_2} = 1.2$ are responsible for asymmetrically scattering
two pencil beams with the tilt angles of $({10^{\circ}},{180^{\circ}})$ and $({50^{\circ}},{270^{\circ}})$. As can be deduced from \textbf{Fig. 11},
although the architectures with two-level quantization (16 distinct phase/amplitude response) fail to
achieve satisfactory results in comparison to those of
continuously modulated designs, the metasurfaces with three-level quantization (64 distinct phase/amplitude response) efficiently operate. Eventually, one can deduce that the superposition operation of terms with unequal coefficients based on the Huygens principle does not remain valid under aggressive quantization ($<$3-bit). 
\begin{figure}[H]
	\centering
	\includegraphics[height=9cm]{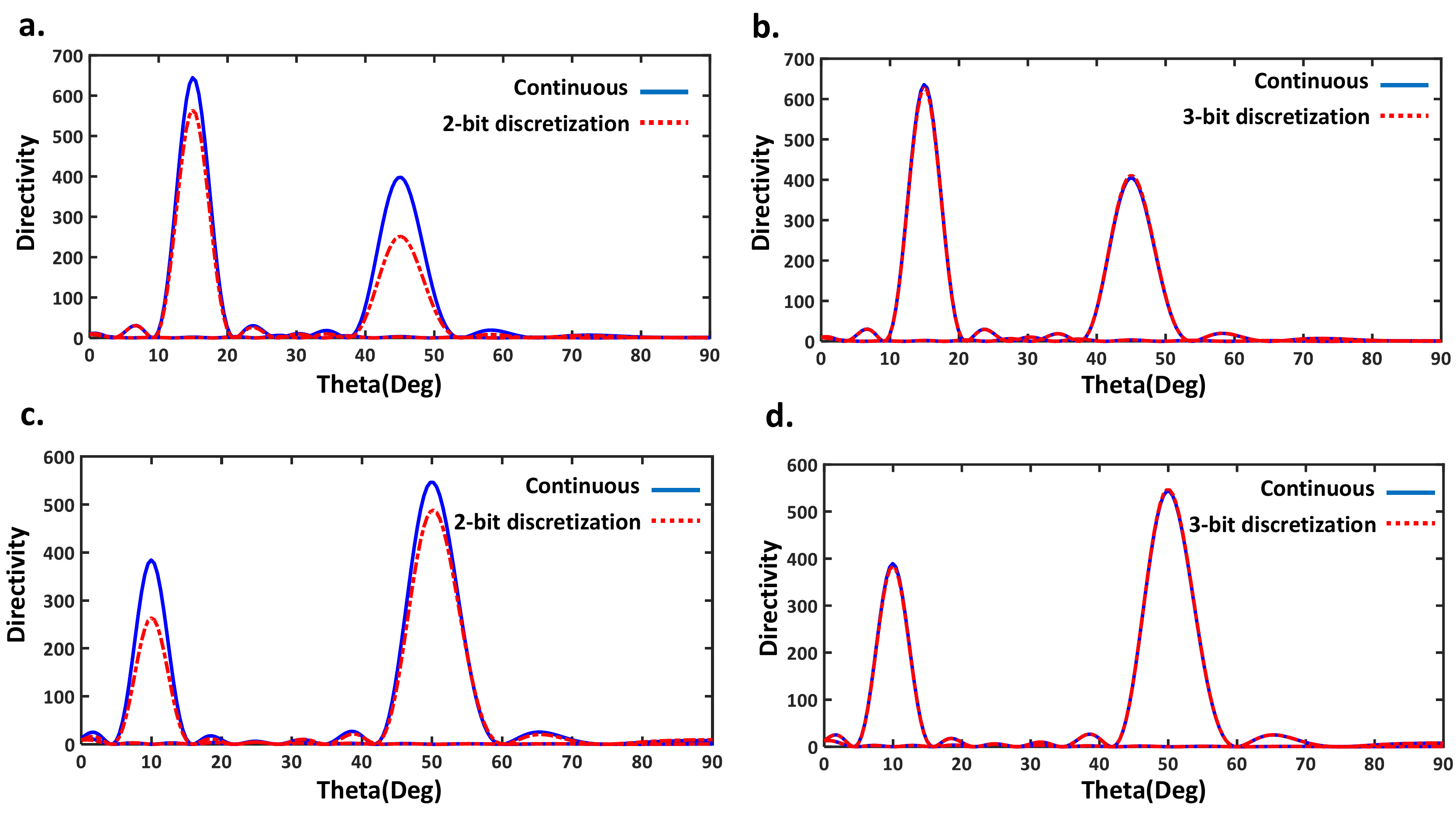}
	\caption{1D directivity intensity pattern (linear format) for space-coding metasurface with asymmetric beams toward $({15^{\circ}},{180^{\circ}})$ and $({45^{\circ}},{270^{\circ}})$ directions assuming (a) two-bit and (b) three-bit quantization. (c,d) 1D directivity intensity pattern (linear format) for space-coding metasurface with asymmetric beams toward $({10^{\circ}},{180^{\circ}})$ and $({50^{\circ}},{270^{\circ}})$ directions assuming two-bit and three-bit quantization.}
	\label{fgr:example2col}
\end{figure}
\section{Conclusion}
In summary, in the first section, benefited from superposition operation with unequal coefficient, Huygens principle and proposed time-varying biasing mechanism, some useful closed-form formulas were presented to manipulate the power intensity pattern in a large space-time digital metasurface utilizing a simple phase-only meta-particle. We applied some simplifying assumptions to reach these convenient formulas and we have demonstrated that these assumptions are valid with a very good approximation in an almost large digital metasurface. Besides, the impact of the metasurface dimensions in the validation of these equations has been addressed. Moreover, the impact of quantization level on the performance of power manipulating has been discussed. The numerical results are in good accordance with analytical predictions and the disorders generated from approximations are very negligible even for scattered beams with large scan angle. In the second section, in the harmonic beam steering scheme, we have presented closed-form formulas to estimate the exact value of absolute directivity at any harmonic frequencies. Utilizing several suitable assumptions, we have derived two separate expressions for calculating the exact total radiated power at harmonic frequencies and total radiated power for scattered beams located at the end-fire direction. The quantitative comparison between simulation results and theoretical predictions have revealed amazing agreement. By observing the introduced limits in the manuscript, the proposed straightforward approach in both sections is expected to broaden the applications of digital space-time metasurfaces significantly and exposes a new opportunity for various applications such as multiple-target radar systems and NOMA communication.  

\section{Competing interests}
The author declare no competing interests.




\end{document}